# Deconvolution of mixing time series on a graph

**Alexander W. Blocker**
Department of Statistics
Harvard University
Cambridge, MA 02138

**Edoardo M. Airoldi**
Department of Statistics
Harvard University
Cambridge, MA 02138

## Abstract

In many applications we are interested in making inference on latent time series from indirect measurements, which are often low-dimensional projections resulting from mixing or aggregation. Positron emission tomography, super-resolution, and network traffic monitoring are some examples. Inference in such settings requires solving a sequence of ill-posed inverse problems, $y_t = Ax_t$, where the projection mechanism provides information on $A$. We consider problems in which $A$ specifies mixing on a graph of times series that are bursty and sparse. We develop a multilevel state-space model for mixing times series and an efficient approach to inference. A simple model is used to calibrate regularization parameters that lead to efficient inference in the multilevel state-space model. We apply this method to the problem of estimating point-to-point traffic flows on a network from aggregate measurements. Our solution outperforms existing methods for this problem, and our two-stage approach suggests an efficient inference strategy for multilevel models of multivariate time series.

## 1 INTRODUCTION

A pervasive challenge in the analysis of multivariate dynamic data is the separation of individual time series from aggregate measurements, known as deconvolution. This flavor of inference problem arises in a number of applications, including super-resolution imaging and positron emission tomography where we want to combine many 2D images in a 3D image consistent with 2D constraints (Shepp and Kruskal, 1978; Vardi et al., 1985); blind source separation where there are more sources than observations (e.g., sound tracks) available (Lee et al., 1999; Parra and Sajda, 2003; Liu and Chen, 1995); and inference on cells of a contingency table where two-way and multi-way margins are given (Bishop et al., 1975; Dobra et al., 2006). The core inference task underlying all of these problems is an ill-posed linear inverse problem $y = Ax$ (Hansen, 1998) where all the components of $y, x$ are non-negative, and $y$ has a lower dimensionality than $x$.

We explore an application of this deconvolution problem known as dynamic network tomography: the problem of inferring origin-destination flows from aggregate traffic on a communication network (Vardi, 1996). It is a classic problem in the statistics and computer science literatures (Vanderbei and Iannone, 1994; Vardi, 1996; Tebaldi and West, 1998; Cao et al., 2000, 2001; Coates et al., 2002; Medina et al., 2002; Zhang et al., 2003; Liang and Yu, 2003; Airoldi and Faloutsos, 2004; Lakhina et al., 2004; Lawrence et al., 2006; Fang et al., 2007). Aggregate traffic flows measured every five minutes, $y_t$, are modeled as a deterministic function, $Ax_t$, of the routing matrix $A$ and the individual origin-destination (OD) flows $x_t, \forall t$. The challenge is that the routing matrix $A$ of size $(r \times c)$ is rank-deficient. In a network with $n$ routers and switches, we typically measure $r = O(n)$ aggregate traffic flows, and we need to infer $c = O(n^2)$ origin-destination flows at five minute intervals. The inference task of interest is filtering, that is, inferring a distribution on $x_t$ given observations $y_{1:t}$ up to time $t$ (estimating $P(x_t|y_1, \ldots, y_t)$). Predicting future OD flows $x_{t+k}$ is a separate, but related, task that relies on accurate filtering.

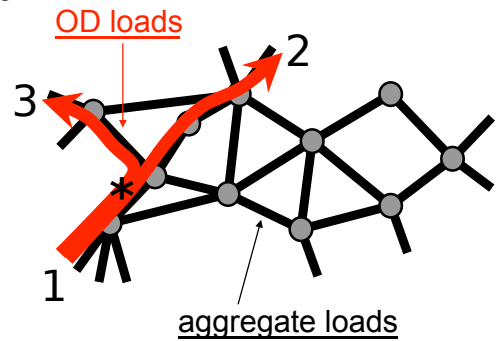

Figure 1: Traffic time-series on a network. The OD loads $1 \to 2$ and $1 \to 3$ are measured as aggregated load $*$.

Figure 1 illustrates this situation. Nodes in the network represent routers and switches. Links between them represent cables connecting them. We want to know the directed OD traffic flows from 1 to 2 and from 1 to 3. However, we can only measure aggregate directed traffic on the links, to which these OD traffic flows both contribute. They cannot be distinguished without additional information.

Here, we develop a multilevel state-space model that posits explicit probabilistic dynamics consisting of two layers. These layers provide a flexible, interpretable structure for inference and capture two vital properties of actual network traffic: spikes in traffic levels and sparsity (an abundance of time periods with small or zero traffic levels). We also develop a two-stage inference strategy for this model. We first fit a simpler, identified Gaussian state-space model to our observed aggregate flows for calibration. We use these estimates to set regularization parameters for the probabilistic dynamics in the multilevel model, developing a proposal for a sequential sample-importance-resample-move (SIRM) particle filter. The sampler is complemented by a new variant of the random directions sampler (Smith, 1984), allowing for efficient inference in a complex dynamic setting. We show that our two-stage inference scheme is more accurate than state-of-the-art methods available, scales to large networks (both in terms of speed of computation and in terms of efficiency of the sequential SIRM sampler), and can be implemented in an online setting.

The success of two-stage strategy relies only on the identifiability of the first-stage model, and can be applied to more general models of ill-posed linear inverse problems. The principle underlying our two-stage strategy can be generalized to other complex multilevel models.

In this sense, our work is related to the recent body of work on "coarse-to-fine" learning techniques (Geman, 2010; Bagnell, 2010; Weis and Taskar, 2010), which include a number of inference strategies for very large data sets, for the most part, but also for complex models of medium size data sets, as in our problem setting. Coarse-to-fine learning strategies typically combine search strategies with inference strategies (Langford, 2010). Our proposed strategy is less focused on search algorithms; rather, it falls into the more statistical traditions of "principled corner cutting" (Meng, 2010) and data-dependent regularization (Clogg et al., 1991).

## 2 A MODEL OF MIXING TIME SERIES ON A GRAPH

Given $m$ observed traffic counters over time, $y_{it}$, the link loads, we want to make inference on $n$ non-observable point-to-point traffic time series, $x_{jt}$, origin-destination traffic flows. The routing scheme is parametrized by the routing matrix $A$, of size $m \times n$. We consider the case of a fixed routing scheme, in which the matrix $A$ has binary entries. Entry $a_{ij}$ specifies whether traffic counter $i$ includes the traffic volume on the origin destination $j$.

We develop a multilevel state-space model to explain the variability in the observed link loads. This data generating process decouples the variability of the non-observable origin destination time series, $x_{1:T}$, into a smooth linear process $\{\lambda_t : t \geq 1\}$ and an independent spike process $\{x_t : t \geq 1\}$. The coefficient that drives the dynamics in the smooth linear process introduces additional variability. Variable dynamics are key for introducing calibrated regularization parameters in the two-stage estimation process.

In detail, we posit that each OD flow $x_{i,t}$ has its own time-varying intensity $\lambda_{i,t}$. This underlying intensity evolves through time according to a multiplicative process

$$\log \lambda_{i,t} = \rho \log \lambda_{i,t-1} + \varepsilon_{i,t}$$

where $\varepsilon_{i,t} \sim \mathrm{N}(\theta_{1\,i,t}, \theta_{2\,i,t})$. This process leads to bursty traffic flows that are not sparse. Moreover, small differences between low traffic flows receive quite different probabilities under this model. Thus, conditional on the underlying intensity, we posit that the latent OD flows $x_{i,t}$ follow a truncated normal error model,

$$x_{i,t} | \lambda_{i,t}, \phi_t \sim \mathrm{TruncN}_{(0,\infty)} \left( \lambda_{i,t},\ \lambda_{i,t}^\tau (\exp(\phi_t) - 1) \right)$$

This error model induces sparsity while maintaining analytical tractability of the inference algorithms (see next Section) by decoupling sparsity control from the bursty dynamic behavior. The parameter $\tau$ provides some flexibility and it can be set with exploratory data analysis on the observed link loads. The mean-variance structure of the error model is analogous to that for a log-normal distribution; in particular, if $\log(z) \sim \mathrm{N}(\mu, \sigma^2)$, $\mathbb{E}(z) = \exp(\mu + \sigma^2/2)$ and $\mathrm{Var}(Z) = \exp(2\mu + \sigma^2) \cdot (\exp(\sigma^2) - 1)$. Thus, $\lambda_{i,t}$ is analogous to $\exp(\mu + \sigma^2/2)$, and $\phi_t$ is analogous to $\sigma^2$. The observed link flows are given by $\boldsymbol{y}_t = A\boldsymbol{x}_t$. We complete our model by placing diffuse independent log-Normal priors on $\lambda_{i,0}$. We also place priors on $\phi_t$ for stability, assuming $\phi_t \sim \mathrm{Gamma}(\alpha, \beta_t/\alpha)$.

The use of this multilevel structure provides a realistic model for the flows were are interested in, which are both bursty and sparse. The log-Normal layer provides bursty dynamics and replicates the intense spikes in traffic observed empirically, whereas the truncated Normal layer allows for very low traffic levels with non-negligible probability. By combining these two distributions, we obtain an overall distribution for our flows that allows for both extreme counts and sparsity, as can be seen in Fig. 2.

The proposed model is capable of generating data that qualitatively resembles our observed flows. This includes the "spike" dynamics observed in the actual flows, as the example in Fig. 3 suggests.

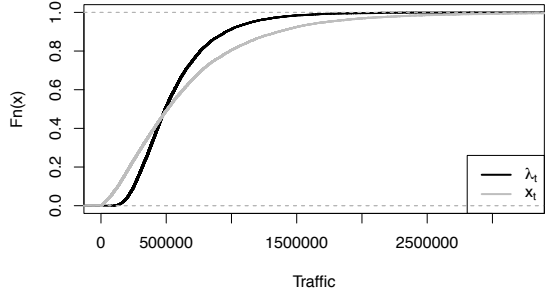

Figure 2: Comparison of CDFs for $\lambda_{i,t}$ and $x_{i,t}$

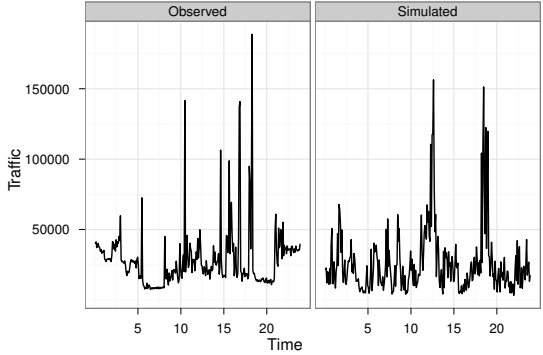

Figure 3: Actual & simulated OD flows

### 2.1 UNIMODALITY & POSTERIOR UPDATES

One important question about this model is how our posterior inferences will behave under dynamic updates. If it tends to "grow" modes over time or exhibit other pathological behavior, our computation would be quite difficult and our inferences would be less credible. Fortunately, this is not the case. In general, we have established that the quasiconcavity of a predictive distribution $f(\boldsymbol{x}_t|\boldsymbol{y}_{t-1},\ldots)$ implies the quasiconcavity of the posterior $f(\boldsymbol{x}_t|\boldsymbol{y}_t,\boldsymbol{y}_{t-1},\ldots)$; thus, the set of maxima for $f(\boldsymbol{x}_t|\boldsymbol{y}_t,\boldsymbol{y}_{t-1},\ldots)$ will form a convex set under the given condition. The proof of this is laid out in Theorem 1:

**Theorem 1.** *Assume $f(\boldsymbol{x}_t)$ is quasiconcave. Let $\boldsymbol{y}_t = A\boldsymbol{x}_t$. Then, $f(\boldsymbol{x}_t|\boldsymbol{y}_t)$ will also be quasiconcave; in particular, it will have no separated modes (the set $\{\boldsymbol{z}: f(\boldsymbol{z}|\boldsymbol{y}_t) = \max_{\boldsymbol{w}} f(\boldsymbol{w}|\boldsymbol{y}_t)\}$ is connected).*

*Proof.* $f(\boldsymbol{x}_t|\boldsymbol{y}_t) \propto I(y_t = A\boldsymbol{x}_t)f(\boldsymbol{x}_t)$, so $f(\boldsymbol{x}_t|\boldsymbol{y}_t)$ has support on only a bounded $r - c$ dimensional subspace of $\mathbb{R}^c$, which forms a closed, bounded, convex polytope in the positive orthant. Denote this region $B(\boldsymbol{y}_t)$. Denote the mode of $f(\boldsymbol{x}_t)$ as $\hat{\boldsymbol{x}}_t$. We now consider two cases:

1. $\hat{\boldsymbol{x}}_t \in B(\boldsymbol{y}_t)$. Then, the mode of $f(\boldsymbol{x}_t|\boldsymbol{y}_t)$ is also $\hat{\boldsymbol{x}}_t$.

2. $\hat{\boldsymbol{x}}_t \notin B(\boldsymbol{y}_t)$. Then, we must be a bit more clever. Consider the level surfaces of $f(\boldsymbol{x}_t|\boldsymbol{y}_t)$, denoting $C(z) = \{\boldsymbol{u}: f(\boldsymbol{u}|\boldsymbol{y}_t) = z\}$. Define $z^* = \max_{B(\boldsymbol{y}_t)} f(\boldsymbol{x}_t|\boldsymbol{y}_t)$; this is well-defined and attained as $B(\boldsymbol{y}_t)$ is closed. Now, denoting $C_0(z) = \{\boldsymbol{u}: f(\boldsymbol{u}) = z\}$, we have $C(z) = C_0(z) \bigcap B(\boldsymbol{y}_t)$. As $f(\boldsymbol{x}_t)$ is quasiconcave, its superlevel sets $U_0(z) = \{\boldsymbol{u}: f(\boldsymbol{u}) \geq z\}$ are convex. Thus, the superlevel sets of $f(\boldsymbol{x}_t|\boldsymbol{y}_t)$, denoted $U(z) = U_0(z) \bigcap B(\boldsymbol{y}_t)$ analogously, are also convex. So, we have that the set $U(z^*) = C(z^*)$ is convex and non-empty. Therefore, we have established that set of modes for $f(\boldsymbol{x}_t|\boldsymbol{y}_t)$ is convex, hence connected.

□

As we initialize our model with a unimodal (quasiconcave) log-Normal distribution and impose log-Normal dynamics on our underlying intensities $\boldsymbol{\lambda}_t$, the above theorem provides a useful limit on pathological behavior for our method.

## 3 INFERENCE

We now outline the inference strategies and computational techniques developed to produce estimates from the previously described model. We use a sequential Monte Carlo algorithm (SIRM) to obtain estimates from our multilevel state-space model, yielding an efficient inference algorithm. The use of a random direction proposal on the feasible region for OD flows is a vital component of this method; without it, sampling these constrained variables would be difficult. However, even with such algorithmic refinements, inference with our model remains difficult without further regularization. Therefore, we use a stable, identifiable calibration model to estimate regularization parameters for our multilevel state-space model. This calibration step is described in detail below.

### 3.1 MODEL-BASED REGULARIZATION

To calibrate regularization parameters for our generative model, we use a simpler model, assuming $\boldsymbol{x}_t$ follows a Gaussian autoregressive process, in Eq 1. This amounts to a standard Gaussian state-space formulation, detailed in Eq. 2. The model we used to obtain the preliminary estimates for the OD flows is:

$$\begin{cases} \boldsymbol{x}_t &= F \cdot \boldsymbol{x}_{t-1} + Q \cdot \boldsymbol{1} + \boldsymbol{e}_t \\ \boldsymbol{y}_t &= A \cdot \boldsymbol{x}_t + \boldsymbol{\epsilon}_t \end{cases} \quad (1)$$

$$= \begin{cases} \begin{bmatrix} \boldsymbol{x}_t \\ \boldsymbol{1} \end{bmatrix} = \begin{bmatrix} F & Q \\ 0 & I \end{bmatrix} \begin{bmatrix} \boldsymbol{x}_{t-1} \\ \boldsymbol{1} \end{bmatrix} + \begin{bmatrix} \boldsymbol{e}_t \\ \boldsymbol{0} \end{bmatrix} \\ \boldsymbol{y}_t = [A|\boldsymbol{0}] \begin{bmatrix} \boldsymbol{x}_t \\ \boldsymbol{1} \end{bmatrix} + \boldsymbol{\epsilon}_t \end{cases}$$

$$= \begin{cases} \tilde{\boldsymbol{x}}_t &= \tilde{F} \cdot \tilde{\boldsymbol{x}}_{t-1} + \tilde{\boldsymbol{e}}_t \\ \boldsymbol{y}_t &= \tilde{A} \cdot \tilde{\boldsymbol{x}}_t + \boldsymbol{\epsilon}_t \end{cases} . \quad (2)$$

We estimate $Q$ and $\text{Cov } e_t$, fixing the remaining parameters. $F$ is fixed at $\rho I$ for simplicity of estimation, with 0.1 a typical value for $\rho$. We also fix $\text{Cov } \epsilon_t$ at $\sigma^2 I$, with 0.01 a typical value for $\sigma^2$. We assume $Q$ to be a positive, diagonal matrix, $Q = \text{diag}(\lambda_t)$, and model $\text{Cov } e_t$ as $\Sigma_t = \phi \, \text{diag}(\lambda_t)^\tau$, where the power is taken entry-wise. We obtain inferences from this model via maximum likelihood on overlapping windows of a fixed length. Implementation details are discussed the next section.

As the marginal likelihood for this model depends only upon the means and covariances of our data, it will be identifiable under conditions analogous to those give in Cao et al. (2000).

**Maximum Likelihood Via Kalman Smoothing.** Maximum likelihood inference for our Gaussian SSM is feasible with standard Kalman smoothing. Two approaches to this maximization problem are possible: EM and direct numerical optimization. The EM approach, outlined for the unconstrained case by Ghahramani and Hinton (1996), requires Kalman smoothing for the E step and maximization of the expected log-likelihood for the M step. The former is straightforward and efficient to calculate using standard algorithms, but the latter requires expensive numerical optimization in our case. Due to the constraints on $Q$ and $\text{Cov } e_t$ and the dependence of our observations, there is no analytic form for the maximum of the expected log-likelihood under our model. Therefore, EM is less favorable; linear convergence is a high price to pay when numerical optimization is required for either approach. We instead use direct numerical optimization on the marginal likelihood obtained from the Kalman smoother, optimizing

$$\ell(Y \mid \theta) = -\sum_t \log |\hat{\Sigma}_t| - \tfrac{1}{2} \sum_t (y_t - \hat{y}_t)' \hat{\Sigma}_t^{-1} (y_t - \hat{y}_t)$$

where $\hat{y}_t$ and $\hat{\Sigma}_t$ are the estimated mean and covariance matrices from the Kalman smoother. With a fast (Fortran) implementation of the Kalman iterations, this approach yields favorable run-times and stable results.

### 3.2 SETTING REGULARIZATION PARAMETERS

To calibrate the regularization parameters for the parameters of the multilevel state-space model, we first run our estimates from the Gaussian SSM at each time through IPFP (the iterative proportional fitting procedure) to ensure positivity and validity with respect to our linear constraints. We then run these estimates through IPFP, ensuring positivity and feasibility for each estimate, and smooth these estimates using a running median with a small window (5 observations in our case), obtaining $\hat{x}_t$. These are used to set $\theta_{1\,i,t}$:

$$\theta_{1\,i,t} = \log \hat{x}_{i,t} - \log \hat{x}_{i,t-1}$$

The variability parameter $\theta_{2\,i,t}$ is set using the estimated variance of each $x_{i,t}$ from our inference with the calibration model. Denoting this estimate as $\hat{V}_{i,t}$, we set $\theta_{2\,i,t}$ as:

$$\theta_{2\,i,t} = (1 - \rho^2) \log(1 + \hat{V}_{i,t}/\hat{x}_{i,t}^2)$$

Here, $\rho$ is the (fixed) autocorrelation parameter in our model for the dynamics of $\log \lambda_{i,t}$. We typically set $\rho = 0.9$. We fix $\rho$ higher for this method than for our Gaussian SSM because more pooling of information across times is necessary. While the Gaussian SSM is identifiable with a sufficiently wide window, our multilevel state-space model relies more strongly on dependence between flows at similar times to obtain information on the underlying parameters and OD flows. Therefore, a larger value amount of autocorrelation is required to obtain stable estimation. Furthermore, a high value for $\rho$ is a practically plausible assumption, as OD flows tend to be highly autocorrelated in communication networks (Cao et al., 2002).

### 3.3 MULTILEVEL STATE-SPACE INFERENCE: SIRM FILTER

Inference in the multilevel state-space model is performed with a sample-resample-move algorithm, akin to Gilks and Berzuini (2001); its structure is outlined in Algorithm 1.

**Sample-Importance-Resample-Move algorithm**
**for** $t \leftarrow t$ **to** $T$ **do**
  Sample step:
    **for** $j \leftarrow 1$ **to** $m$ **do**
      Draw a proposal
      $\log \lambda_{i,t}^{(j)*} \sim \text{N}(\theta_{1\,i,t} + \log \lambda_{i,t-1}^{(j)}, \theta_{1\,i,t})$
      Draw $\phi_t^{(j)} \sim \text{Gamma}(\alpha, \hat{\phi}_t/\alpha)$
      Draw $x_t^{(j)*}$ from a truncated Normal distribution
      with mean $\mu^* = \theta_{1\,t} + \rho/m \sum_{j=1}^m \lambda_{t-1}^{(j)}$ and
      covariance matrix $\Sigma^* = (\exp(\hat{\phi}_t) - 1) \text{diag}(\mu^{*2})$
      on the feasible region given by $x_t^{(j)*} \geq 0$,
      $y_t = A x_t^{(j)*}$ using Algorithm 2
    Resample our particles $(\lambda_t^{(j)*}, \phi_t^{(j)*}, x_t^{(j)*})$ with
    probabilities proportional to our weights $w_t^{(j)}$
    Move each of our resampled particles
    $(\lambda_t^{(j)}, \phi_t^{(j)}, x_t^{(j)})$ using a MCMC algorithm
    (Metropolis-Hastings within Gibbs, with proposal on
    $x_t$ given by Algorithm 2)
**return** $(\lambda_t^{(j)}, \phi_t^{(j)}, x_t^{(j)})$ for $j \leftarrow 1$ **to** $m$, $t \leftarrow 1$ **to** $T$

**Algorithm 1**: SIRM algorithm for inference with multilevel state-space model

We use the approach of Smith (1984), known as the "random directions algorithm" (RDA), to sample from distributions on constrained regions in our algorithm. This method constructs a random-walk proposal on convex regions (such as our feasible regions for $x_t$) by first drawing a vector $d$ uniformly on the unit sphere. We then calculate the intersections of a line along this vector with the surface of the bounding region and sampling uniformly along the feasible segment of this line. This feasible segment can be calculated easily using the decomposition of $A$ given by Tebaldi and West (1998). They decompose $A$ as $[A_1\ A_2]$ by permuting the columns of $A$ (and the corresponding components of $\boldsymbol{x}_t$, where $A_1$ ($r \times r$) is of full rank. Then, splitting the permuted vector $\boldsymbol{x}_t = [\boldsymbol{x}'_{1,t}\ \boldsymbol{x}'_{2,t}]$, we obtain $\boldsymbol{x}_{1,t} = A_1^{-1}(\boldsymbol{y}_t - A_2 \boldsymbol{x}_{2,t})$. This formulation can be used to construct an efficient random directions algorithm to propose valid values of $\boldsymbol{x}_t$; we have included pseudocode for this algorithm in Algorithm 2.

**Random Directions Algorithm**
**Initialization**
  **begin**
    | Decompose $A$ into $[A_1\ A_2]$, $A_1$ ($r \times r$) full-rank as in Tebaldi and West (1998)
    | Store $B := A_1^{-1}$; $C := A_1^{-1} A_2$
  **end**
**Metropolis step**
  **given** $\boldsymbol{x}_t$
  **begin**
    | Draw $\boldsymbol{z} \sim N(0, I)$, $\boldsymbol{z} \in \mathbb{R}^{c-r}$
    | Set $\boldsymbol{d} := \boldsymbol{z}/\|\boldsymbol{z}\|$
    | Calculate $\boldsymbol{w} := C \cdot d$
    | Set $h_1 := \max\{\min_{k:w_k>0}(\boldsymbol{x}_{1,t})_k/w_k, 0\}$
    | Set $h_2 := \max\{\min_{k:d_k<0} -(\boldsymbol{x}_{2,t})_k/d_k, 0\}$
    | Set $h := \min\{h_1, h_2\}$
    | Set $l_1 := \max\{\max_{k:w_k<0}(\boldsymbol{x}_{1,t})_k/w_k, 0\}$
    | Set $l_2 := \max\{\max_{k:d_k>0} -(\boldsymbol{x}_{2,t})_k/d_k, 0\}$
    | Set $l := \max\{l_1, l_2\}$
    | Draw $u \sim \text{Unif}(l, h)$
    | Set $\boldsymbol{x}^*_{2,t} := \boldsymbol{x}_{2,t} + u \cdot \boldsymbol{d}$; $\boldsymbol{x}^*_{1,t} = \boldsymbol{x}_{1,t} - u \cdot w$;
    | $\boldsymbol{x}^*_t = (\boldsymbol{x}^*_{1,t}, \boldsymbol{x}^*_{2,t})$
    | Set $\boldsymbol{x}_t := \boldsymbol{x}^*_t$ with probability $\min\{f(\boldsymbol{x}^*_t)/f(\boldsymbol{x}_t), 1\}$
  **end**
  **return** $\boldsymbol{x}_t$

**Algorithm 2**: RDA algorithm for sampling from $f(\boldsymbol{x}_t)$, truncated to the feasible region given by $A \cdot \boldsymbol{x}_t = \boldsymbol{y}_t$

All draws from this proposal have positive posterior density (as they are feasible). This property allows our sampling methods to move away from problematic boundary regions of the given feasible polytope; methods that use, for example, Gaussian random-walk proposal rules can perform quite poorly in these situations, requiring an extremely large number of draws to obtain feasible proposals. For example, with $\boldsymbol{x}_t \in \mathbb{R}^{16}$, it can sometimes require $10^9$ or more draws to obtain a valid particle using the conditional posterior from $t-1$ as a proposal with the datasets tested in Section 5.

## 4 SIMULATION STUDY

To evaluate our inference algorithm, we simulated OD flows from our multilevel state-space model under 3 network topographies: a 3-node bidirectional chain, a 3-node star topography, and a 4-node start topography, corresponding to 2, 4, and 9-dimensional latent spaces for our inference on $\boldsymbol{x}_t$. For each of these cases, we produced 30 replicates consisting of 300 time-points. We then calculated the link flows for each replicate and ran our inference algorithm. In addition to the two-stage approach outlined previously, we also performed filtering using our multilevel state-space model with a naïve random-walk regularization on the OD flows; that is, we set $\theta_{1\,i,t} = 0\ \ \forall (i,t)$ and $\theta_{2\,i,t} = \log(5)/2$. This allows us to directly evaluate the effect of our regularization strategy and plausibility of our model.

The primary quantity of interest in our simulations are the relative mean $L_2$ and $L_1$ errors in estimated OD flows for for the naïve regularization compared to our two-stage method. The distributions of these relative errors is summarized in Figure 4.

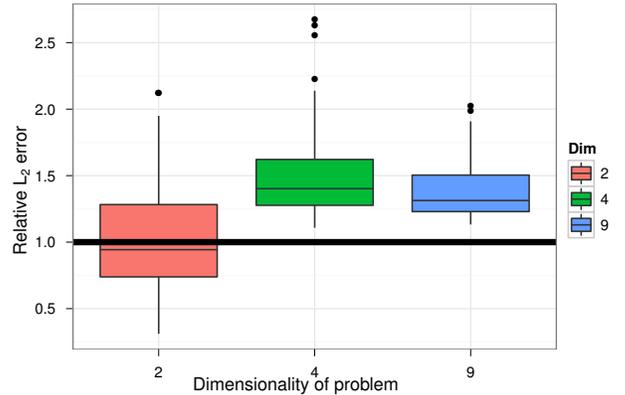

Figure 4: Relative $L_2$ error for naïve vs. two-stage method against dimensionality

We find that our two-stage method outperforms one using a naïve regularization. The decreasing variation in our relative errors suggests that the consistency of this outperformance increases with dimensionality. Specifically, we have a mean relative error of $1.09 \pm 0.49$ in 2 dimensions, increasing to $1.57 \pm 0.45$ in 4 dimensions and $1.40 \pm 0.26$ in 9 dimensions. Our experience with these simulations also highlighted the computational benefits of our two-stage strategy. During particle filtering iterations with the naïve regularization, the effective number of particles rarely climbed above 2, whereas we typically obtained $10-50$ with the two-stage approach (with an equivalent number of particles). With real data, we expect additional benefits from our two-stage approach; in particu-lar, we would expect it to have greater robustness to model misspecification. We are using information from a simpler model to rein-in potential issues with more the more delicate hierarchical model, which should stabilize inferences from the latter and limit problems of non-identifiability. These expectations are borne out in Section 5.

## 5 EMPIRICAL RESULTS

We present two datasets from the field of network tomography, one spanning 4 nodes (16 OD flows) and the other spanning 12 nodes (144 OD flows), upon which we evaluate our two-stage deconvolution method. We compare the performance of our approach to that of several previously presented in the literature for this problem, focusing on accuracy, computational stability, and scalability.

### 5.1 DATASETS

Our first dataset is that of Cao et al. (2000). We use their "Router1" network, consisting of 4 nodes in a star topography (yielding 8 observed link flows) with 16 OD flows. The data consists of link flows observed every 5 minutes over 1 day on a Bell Labs corporate network. This yields 287 times, providing a rich dataset for investigation of the use of dynamic information in network tomography. The small size of this network allows us to focus on the fundamentals of the problem without focusing on issues of scalability.

For further investigation, we constructed a dataset from 2 days of observed OD flows on the CMU network. The routing table for this network is sensitive (due to network security issues), so we combined actual OD flows in a synthetic network topography. This network topology consists of 12 nodes. These are connected in a star topography to two routers (one with 4 of the nodes, the other with the remaining 8); the routers are linked via a single connection. This configuration yields 26 observed link flows and 144 OD flows, observed over 473 times (5 minute intervals). This larger dataset allows for the comparison of network tomography techniques in a richer, more realistic setting. In combination with the "Router1" data, it also allows us to explore the effect of dimensionality on performance and computational efficiency.

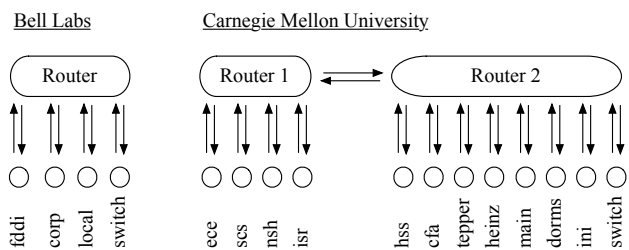

Figure 5: Bell Labs and CMU network structures.

We do not apply any seasonal adjustment or other more complex dynamic models to these datasets. We would recommend such an extension for data spanning longer periods; indeed, even for data spanning only two days, usage patterns by time-of-day can be significant. However, we endeavour to compare our deconvolution algorithms on equal footing – our focus is dynamic deconvolution. Thus, all methods are implemented with only local dynamics (no seasonal adjustment).

### 5.2 COMPETING METHODS

We tested the locally IID and smoothed methods of Cao et al. (2000), the Bayesian MCMC approach of Tebaldi and West (1998), our multilevel state-space model with naïve regularization (as in Section 4), and our two-stage approach (Gaussian SSM followed by multilevel state-space model). All approaches were implemented in R with extensions in C for particular bottlenecks (e.g. IPFP). For the methods of Cao et al. (2000) and our Gaussian SSM, which use windowed estimates, we used a window width of 23.

For the approach of Tebaldi and West (1998), we tested both the original implementation and our own modification in which (following the authors' original notation) $\lambda_j$ and $X_j$ are sampled with a joint Metropolis-Hastings step. The proposal distribution for this step is constructed by first proposing uniformly along the range of feasible values for $X_j$ given all other values, then drawing $\lambda_j$ from its conditional posterior given the proposed $X_j$. This greatly improves the efficiency of inference for their model, leading to improved convergence (we observed multivariate Gelman-Rubin diagnostics reduced by approximately an order of magnitude) and better predictions. These improvements allow us to compare our model to theirs on a more level playing field, focusing on the underlying model rather than computational issues.

### 5.3 PERFORMANCE COMPARISON

We summarize performance of the previously mentioned methods on both datasets in Table 1. Each row corresponds to a method, and the columns provide average $L_1$ and $L_2$ errors for the estimates of OD flows in each dataset with corresponding standard errors. For the Bell Labs dataset, we provide errors in kilobytes; for the CMU data, we provide errors in megabytes. We also provide Figures S1-S5 in our supplemental material as a visualization of our results on the Bell Labs dataset. We compare performance in accuracy, computational stability, and scalability.

**Accuracy.** We obtain favorable performance for our two-stage approach (corresponding the final row of Table 1) for both datasets. Mean $L_1$ and $L_2$ errors for this method are within 1 SE of the minimum for the Bell Labs data. Both of our methods reduce average $L_1$ and $L_2$ errors by 60-80% compared to the other approaches presented, representing a major gain in predictive accuracy. For the CMU data, we obtain a reduction of 53% in average $L_2$ error and 44% in average $L_1$ error from the algorithm of Tebaldi and West (1998) to our multilevel state-space model; we observe 14-15% reductions in average $L_1$ and $L_2$ errors from our Gaussian SSM to the multilevel state-space models. Furthermore, we observe large gains in filtering performance for both datasets compared to inference using naïve regularization with our multilevel state-space model. Overall, our approach outperforms existing methods in accuracy, with greater gains from the Gaussian SSM to the multilevel state-space model in our higher-dimensional setting.

Table 1: Performance Comparison (Bell Labs results in Kb, CMU results in Mb. *Denotes our own improvement on the original algorithm by Tebaldi & West)

| Method | BELL LABS | | | | CMU | | | |
|---|---|---|---|---|---|---|---|---|
| | $L_2$ Error | SE | $L_1$ Error | SE | $L_2$ Error | SE | $L_1$ Error | SE |
| Locally IID model | 104.59 | 5.54 | 160.24 | 6.53 | 592.49 | 9.91 | 1169.15 | 17.11 |
| Smoothed locally IID | 104.25 | 5.52 | 157.87 | 6.48 | — | — | — | — |
| Tebaldi & West (uniform prior) | 76.60 | 4.91 | 173.94 | 7.49 | — | — | — | — |
| Tebaldi & West (joint proposal)* | 49.43 | 2.58 | 147.66 | 6.18 | 167.94 | 4.42 | 712.37 | 14.68 |
| Dynamic multilevel model (naïve) | 63.29 | 3.35 | 178.43 | 8.09 | 311.21 | 6.25 | 1109.68 | 19.58 |
| Calibration model (stage 1) | 19.35 | 0.72 | 57.66 | 2.06 | 110.47 | 6.19 | 389.14 | 16.72 |
| Dynamic multilevel model (stage 2) | 19.93 | 0.87 | 58.20 | 2.39 | 93.42 | 5.20 | 334.74 | 13.64 |

The three last methods in Table 1 each contain a mix of three attributes: explicit dynamics, skewness, and regularization to improve identifiability. Our multilevel state-space model with naïve regularization incorporates the former two, but its performance suffers from problems with identifiability. Our calibration model is identifiable and incorporates explicit dynamics, but does not account for skewness. It performs well on the Bell Labs dataset, where that distributions of OD flows are relatively symmetric, but suffers on the extremely skewed CMU dataset. Our two-stage procedure using the multilevel state-space model overcomes identifiability issues and accounts for skewness, attaining comparable performance with the Bell Labs network and outperforming considerably on the CMU network as a result.

**Computational stability.** The methods tested varied in computational stability. Those of Cao et al. (2000) remained stable across both datasets, but the original method of Tebaldi and West (1998) encountered issues. On the Bell Labs data, it required a very large number of iterations to obtain convergence (as indicated by the Gelman-Rubin diagnostic); 150,000 iterations per time were used to provide the given estimates, 50,000 of which were discarded as burn-in. This method failed completely on several times in the CMU data, becoming trapped in a corner of the feasible region. Our revised version of their algorithm performed much better, requiring far fewer iterations for convergence (50,000 was sufficient for all examined cases, but 150,000 were used for the results presented for comparability).

Our calibration model proved computationally stable across both datasets. The direct use of marginal likelihood, for maximum likelihood estimation, proved effective even in a high-dimensional setting. The multilevel state-space model was also stable in both settings with the given structure; however, it was sensitive to some of the points mentioned in Section 3. Major problems arose in experiments using the posterior on $x_t$ from the previous time as a proposal (as is common in applications of particle filtering); several times in the Bell Labs data required over 10 million proposals to obtain a single feasible particle. Additional care was needed with the "move" step due to similar issues. Furthermore, the use of a naïve, random-walk regularization caused some computational difficulties, as the particles often became extremely diffuse in the feasible region. Overall, we found inference with the multilevel state-space model computationally stable so long as sampling methods for highly constrained variables ($x_t$ in particular) explicitly respected said constraints, proposing only valid values. Our random directions algorithm (Algorithm 2) handles this task well.

**Scalability.** All methods evaluated fared well in scalability, including our computationally-intensive, simulation-based inference using the multilevel state-space model. We focus first on the run-times for the CMU dataset. For each time, the methods of Cao et al. (2000) required approximately 225 seconds to obtain maximum likelihood estimates with a 23 observation window. Our modification of Tebaldi and West (1998) required approximately 1500 seconds to obtain 150,000 samples for a single time; the original (where it ran) required 2250 seconds on average. In contrast, our simulation-based filtering method for the multilevel state-space model required 270 seconds per time on average. For the Bell Labs dataset, our filtering method required approximately 8 seconds per time, whereas our modification of Tebaldi and West (1998) required 150 seconds per time and their original algorithm required approximately the same. These results are encouraging: the fil-tering component of our method, which would be the part used in online applications, is reasonably efficient (even written in R) and could run in faster than real-time with 144 OD flows at 5 minute sampling intervals. Given more efficient implementation and parallelization (which is feasible for all sampling steps), this approach can scale to problems of the scale typically found in the real-world. This is especially true given the sparsity of many such flows on networks; the prevalence of zero aggregate (link) flows in real-world data reduces the effective size of the deconvolution problem.

# 6 REMARKS

We have addressed the deconvolution of time-series on a graph, with an application to dynamic network tomography. For this problem, we develop a novel statistical machine learning approach to inference by combining a novel two-stage strategy with a new multilevel state-space model that posits non-Gaussian marginals and nonlinear probabilistic dynamics. Our results and analyses substantiate several claims and suggest points for further discussion.

To demonstrate our method, we analyzed two networks, at Bell Labs and CMU, which span substantial range of dimensionality, with different inference methods. The results demonstrate a clear improvement of the proposed methodology over previously published methods in reconstructing OD flows in two network tomography settings. Comparison between Bell Labs and CMU results suggests that this gain increases with the dimensionality of the problem.

## 6.1 MODELING AND COMPUTATIONAL ISSUES

Our model explicitly captures two critical feature of our time series — namely, skewness and temporal dependence. A large portion of the substantial improvements in accuracy over existing methods can be attributed to these modeling improvements. The gains in computational efficiency account for only part of the improvements in accuracy — see below. Previous approaches have included skewness (Tebaldi and West, 1998), but never explicit temporal dependence of the OD flows. The inter-temporal smoothing algorithm of Cao et al. (2000) includes elements of temporal dependence; however, the model assumed temporally independent OD flows and the dependence is on the width of the windows of observations that contribute to inference of the OD flows for each time point. In summary, previous work has not accounted for the range of properties addressed here. The performance gains from such modeling are clear in the two datasets tested; in particular, the gains from the model of Tebaldi and West (1998) to the Gaussian SSM and the final multilevel state-space model for the CMU data reinforce the benefits of using a realistic model in this problem.

Fundamentally, we estimate the OD flows by projecting our observations onto the latent space the flows inhabit; that is, we want to compute $\mathbb{E}[\boldsymbol{x}_t \mid \boldsymbol{y}_t]$ under a given probabilistic structure. The relative variability of OD flows over time plays a large role in inference, as their is typically a strong relationship between the mean and variance of OD flows. Because of this, simple methods that do not model variability explicitly and realistically, including Moore-Penrose generalized inverse (Harville, 2008), independent component analysis (Hyvärinen et al., 2003), and iterative proportional fitting procedure (Fienberg, 1970) are of limited use in this context. Our approach, in contrast, models this variability with a probabilistic structure, improving inferences by using this additional information.

Our inference method is computationally efficient and scales to large networks than have been previously addressed using probabilistic models. The problem is fundamentally $O(c)$ for each time observed, so we cannot hope to do better than quadratic scaling in the number of nodes in our network (excepting cases where link loads are 0). Despite the sophistication of our multilevel state-space model, our sequential Monte Carlo technique allows for inference in better than real-time for a network with 144 OD flows. As this is the component that would be used in an online application, we have demonstrated a scalable technique for inference with a model of greater complexity and realism than has been previously found in the literature.

These gains in computational efficiency also reduce numerical instability and are ultimately responsible for additional gains in accuracy. Computational issues can be appreciated by considering the amount of effort we needed to place in maintaining $\mathrm{Cov}\,\boldsymbol{e}_t$ positive-definite in the EM algorithm of Cao et al. (2000), especially when the traffic approaches zero. We can see some artifacts in the corresponding OD estimates in Figure S1 (green lines) due to this issue in the low traffic OD flows, e.g. "orig local, dest local". We further quantified the effects of computational efficiency on inference in the original methods by Tebaldi and West (1998) in Table 1 by comparing the uniform prior and component-wise proposal to the joint proposal we developed. In addition to the gains in speed and convergence discussed in Section 5.3, we observe a large reduction in average error from the component-wise to joint proposal (35% in $L_2$ error, 15% in $L_1$) with no change in priors or the underlying model.

The use of the random directions algorithm (of Smith (1984)) in our sequential Monte Carlo method is vital. Without such an algorithm to sample directly on the feasible region for each set of OD flows, we would be forced to use a naive proposal distribution. In our testing, such distributions proved extremely problematic (as discussed in Section 3), especially in higher dimensional settings. In such cases, sampling strategies that fully utilize the available constraint information are necessary to obtain high ac-curacy and efficiency. This is particularly salient comparing the results presented here to our previous work (Airoldi and Faloutsos, 2004); the method presented in that work suffered from computational instabilities, requiring restarts of its filtering algorithm at particular time points. It was also hampered by inefficient sampling on the feasible region and distributional assumptions (the log-Normal was used as the distribution of $x_t|\lambda_t$) that induced issues in modeling OD flows near zero. Recently proposed sampling strategies (Airoldi and Haas, 2011) will likely improve computational performance and the estimates.

Last, multi-modality in the marginal posterior on each OD flow $x_{it}$ appears low to non-existent in our investigations. However Tebaldi and West (1998) and Airoldi and Faloutsos (2004) have observed a substantial amount of multi-modality in their results. Our results suggest that the theoretical amount of multi-modality in these problems is low for the case of real-valued OD flows and models assuming independent OD flows. The amount of multi-modality observed in previous work appears to originate primarily from the inefficiency of the samplers and, to a lesser extent, the assumption of discrete-valued OD flows. This further reinforces the importance of efficient computation for inference in complex, weakly identified settings; even a simple model can falter on poor computation, and complex models require great computational care to obtain reliable inferences.

### 6.2 A NOVEL TWO-STATE STRATEGY FOR INFERENCE IN DYNAMIC MULTILEVEL MODELS

As previously argued by Tebaldi and West (1998) in the static setting, informative priors are essential to identify a unique posterior mode that approximates the true configuration of OD flows well. This is remains true in the dynamic setting, despite the additional information that temporal dependence makes relevant for the inference of each OD flow. The technical choices at issue are: (i) where to find such information that it is not obvious in the data; (ii) what parameters are most convenient to put priors on; and (iii) how do we translate the additional information into prior information for the chosen parameters.

We use a simple nearly identifiable model, which is not as realistic as our multilevel state-space model, to find rough estimates of the OD flows (in our first-stage). These estimates provide some information about where the OD flows are in the space of feasible solutions, enabling us to identify high-probability subsets of the feasible region at each time before embarking on computationally-intensive simulations. This effect is larger in higher dimensions, as the proportion of the feasible region's volume with high posterior density decreases rapidly with dimensionality (the classical curse of dimensionality). Practically, informative priors increase the efficiency of the particle filter by focusing its sampling on promising regions of the parameter space, avoiding wasted computation and improving inferences.

In order to pass the first-stage information to the (non-Gaussian) multilevel state-space model, we moved away from a standard linear state-space formulation with additive error to a non-linear formulation with stochastic dynamics, which effectively provides a multiplicative error (second-stage). The stochastic dynamics assumed for $\lambda_t$ provide our parameters of choice for the incorporation of this information; adding this layer of variability to our model allows our regularization to guide inferences without placing too tight a constraint on the inferred OD flows. In Section 3 we described the problem we solve to translate the first-stage estimates into regularization for the parameters of the second-stage model. We traded off the need to pass as much of this information as possible from the first- to second-stage with the known inaccuracy of the first-stage.

Our two-stage procedure suggests a more general strategy for inference in complex hierarchical models with mild identifiability issues. Using simpler models to guide computation with more sophisticated, realistic models and (if necessary) provide regularization can provide large gains in performance. We have demonstrated the utility of this principled approach to "cutting corners". This approach has allowed us to use a sophisticated generative model for inference, leveraging the power of multilevel analysis, while maintaining efficiency for real-time applications. We look forward to investigating its utility in other settings, including predicting flows of communications within social networks from aggregate measurements and inferring biological fluxes from experimental data.

# A APPENDIX – SUPPLEMENTARY MATERIAL

In this appendix, we show the actual vs. fitted OD flows for the methods presented previously. We plot all OD flows for the Bell Labs data and the 12 most variable OD flows for CMU. Ground truth is always in black, with estimated values in color. Figures S1 through S5 cover the Bell Labs data, and Figures S6 through S10 cover the CMU data.

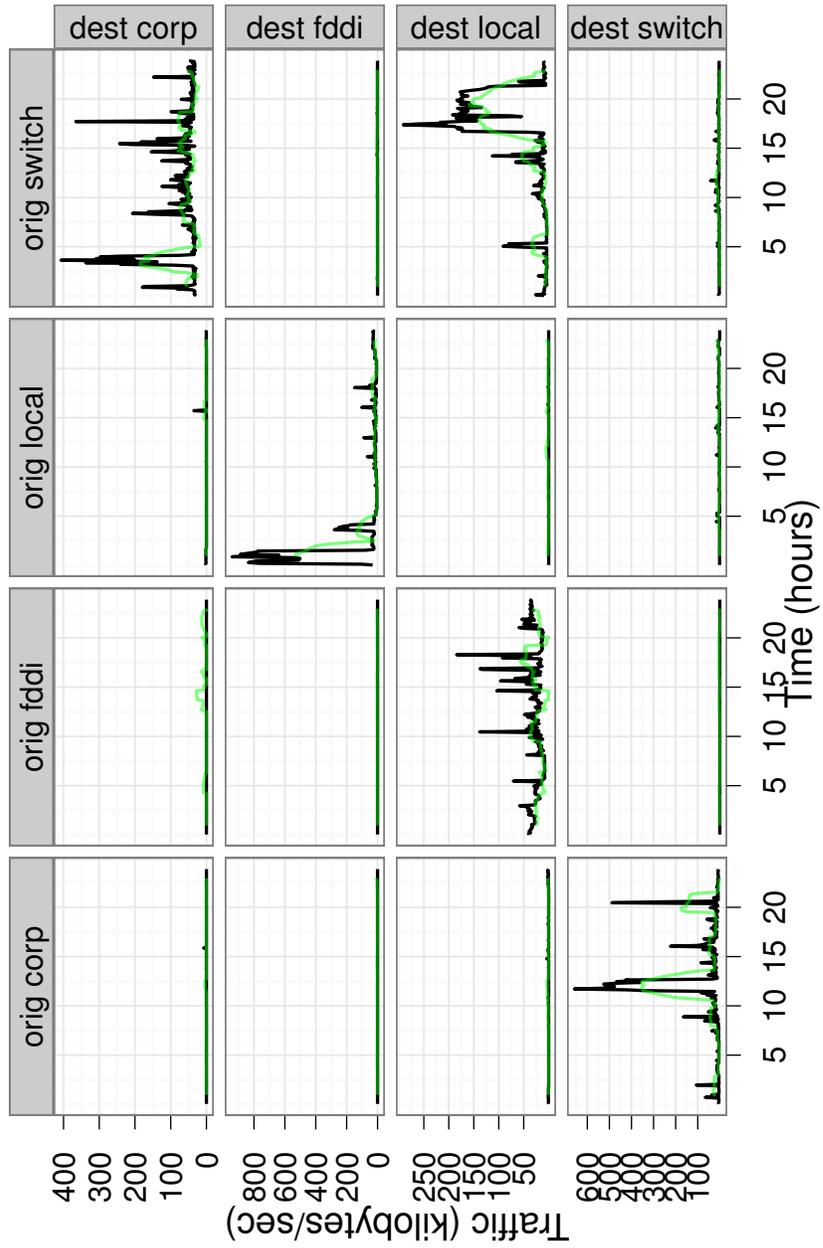

Figure S1: Fitted values vs. ground truth for Bell Labs data. Ground truth in black; Locally IID model in green.

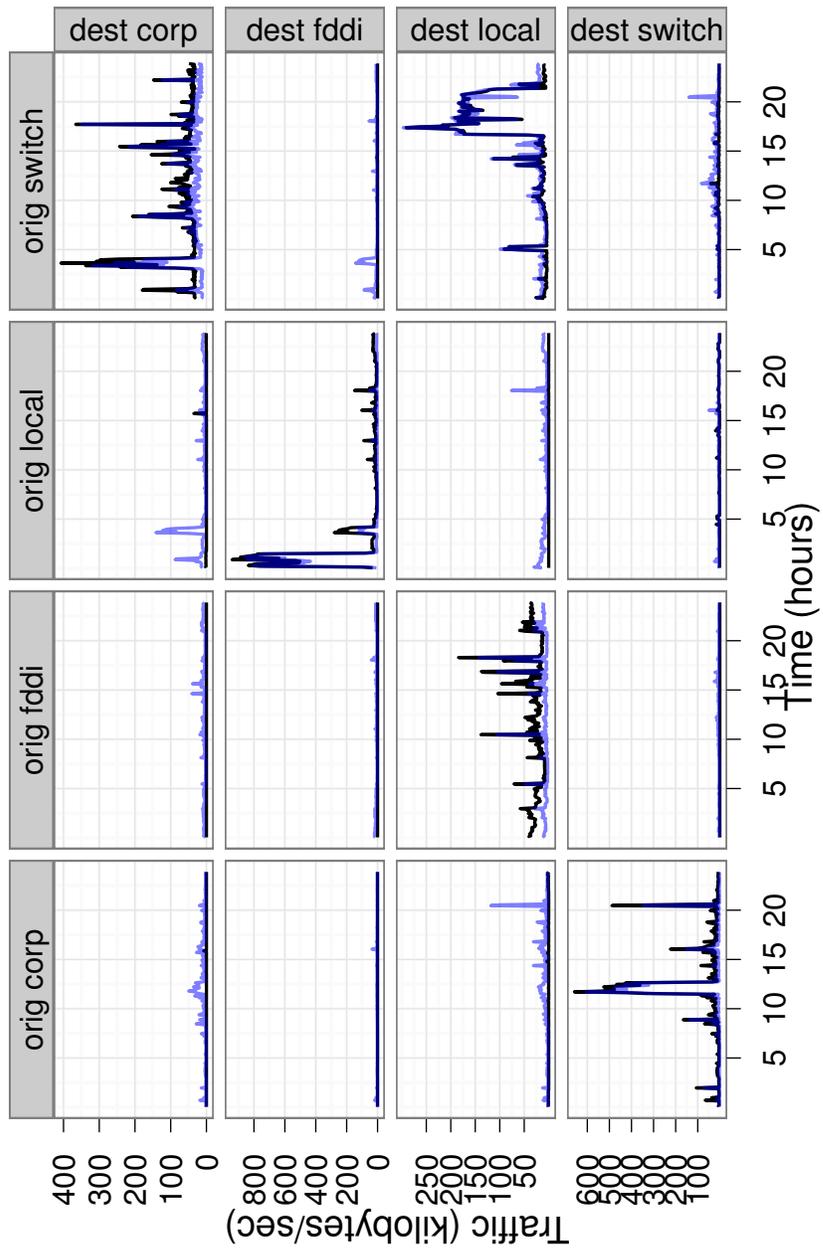

Figure S2: Fitted values vs. ground truth for Bell Labs data. Ground truth in black; Tebaldi & West (joint proposal) in blue.

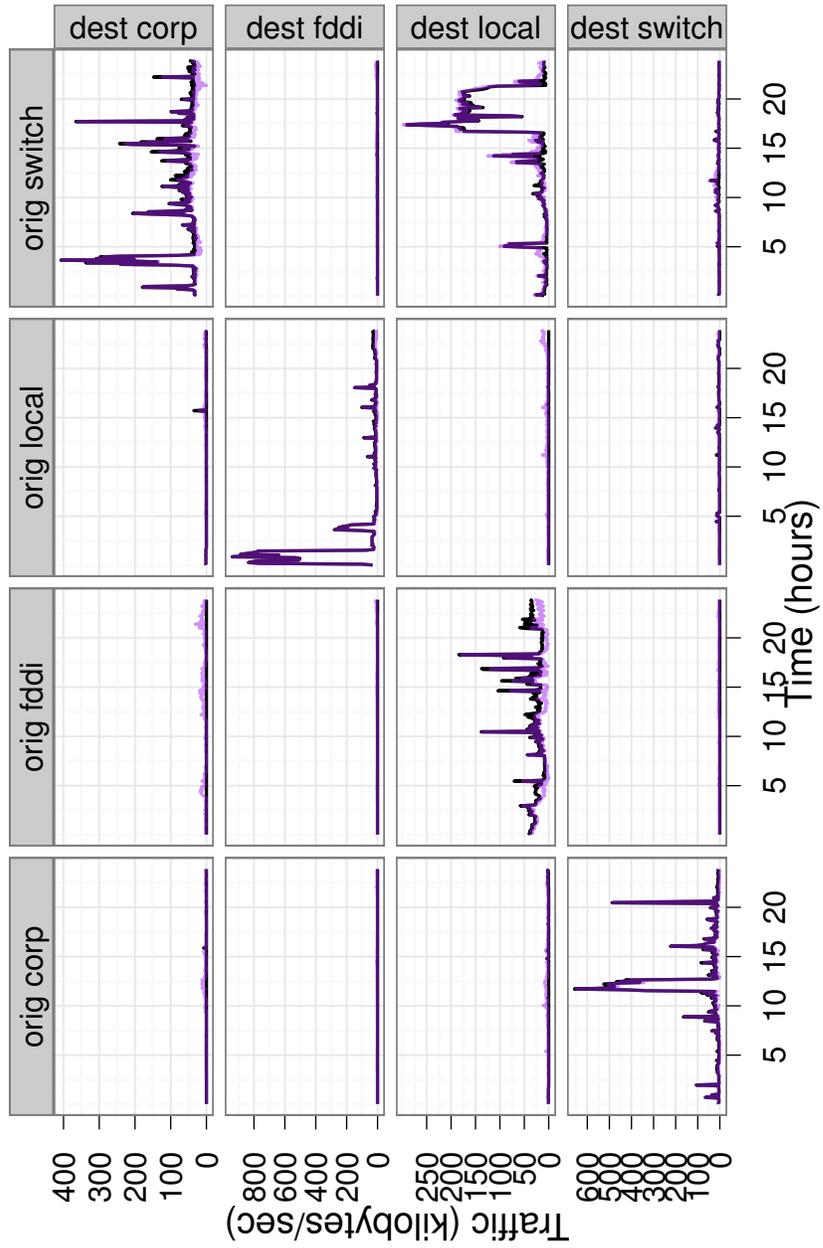

Figure S3: Fitted values vs. ground truth for Bell Labs data. Ground truth in black; Calibration model (stage 1) in purple.

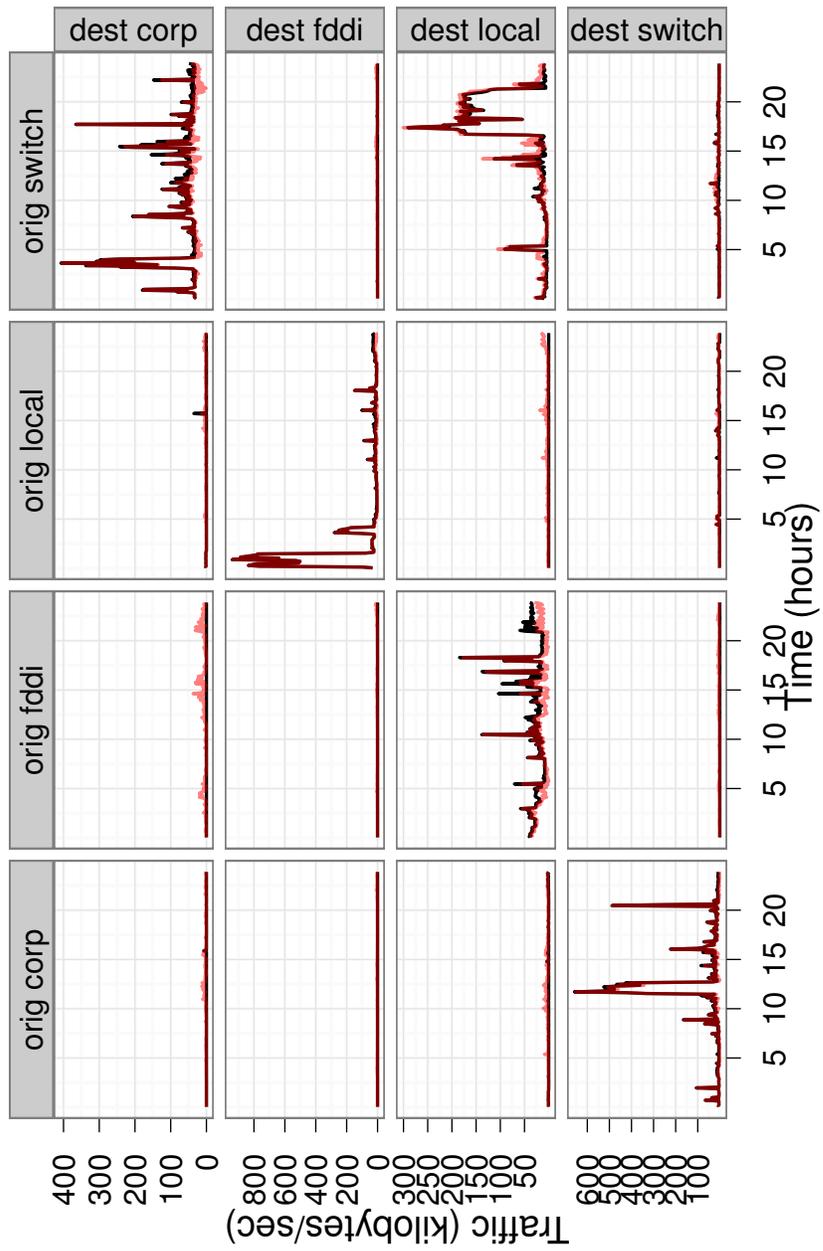

Figure S4: Fitted values vs. ground truth for Bell Labs data. Ground truth in black; Dynamic multilevel model (stage 2) in red.

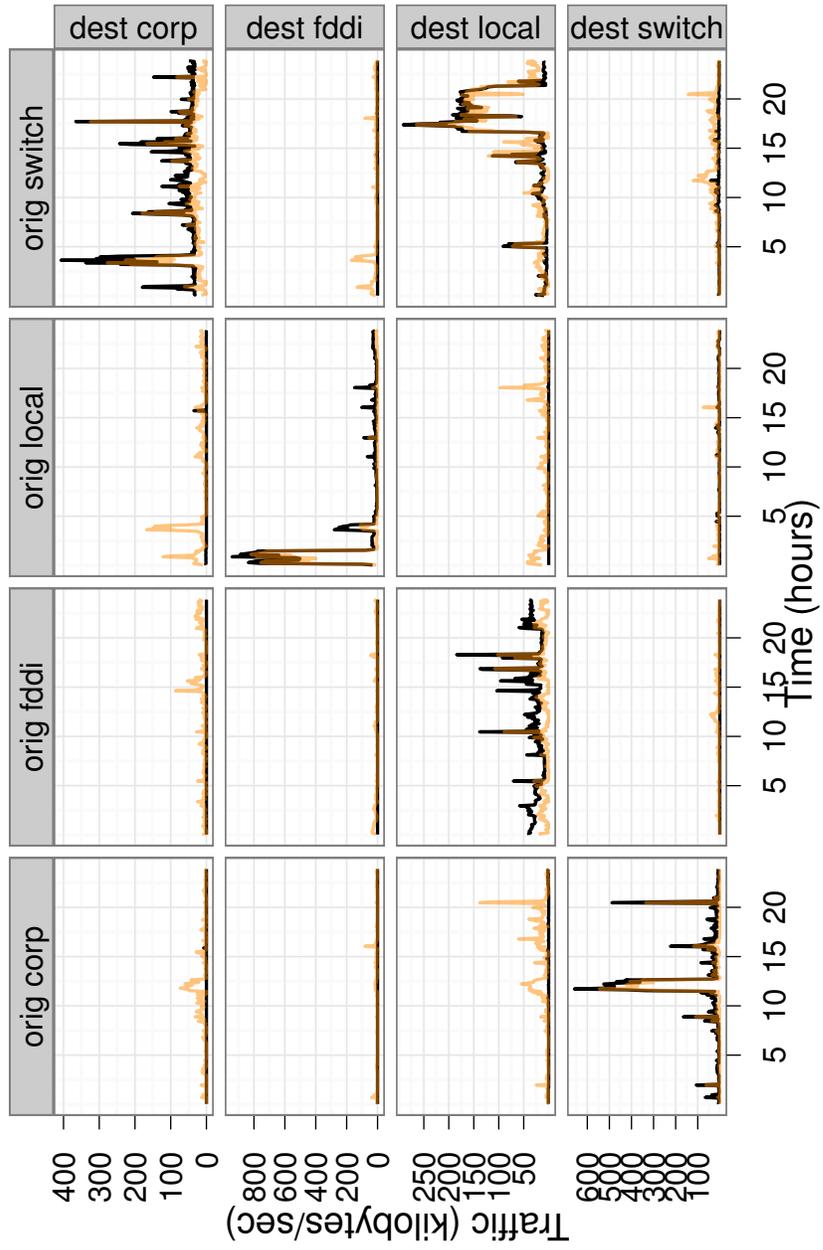

Figure S5: Fitted values vs. ground truth for Bell Labs data. Ground truth in black; Naïve prior in orange.

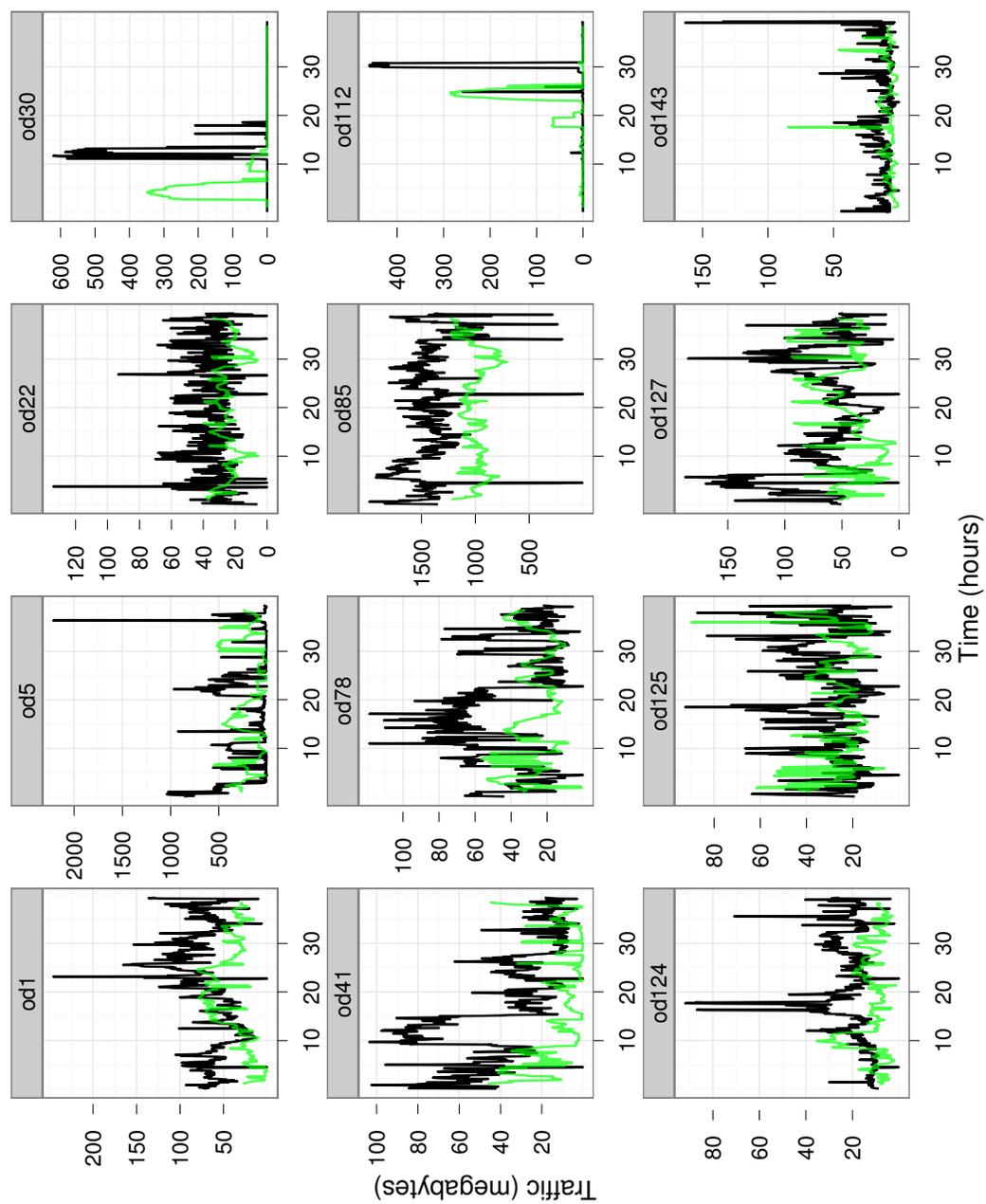

Figure S6: Fitted values vs. ground truth for CMU data. Ground truth in black; Locally IID model in green.

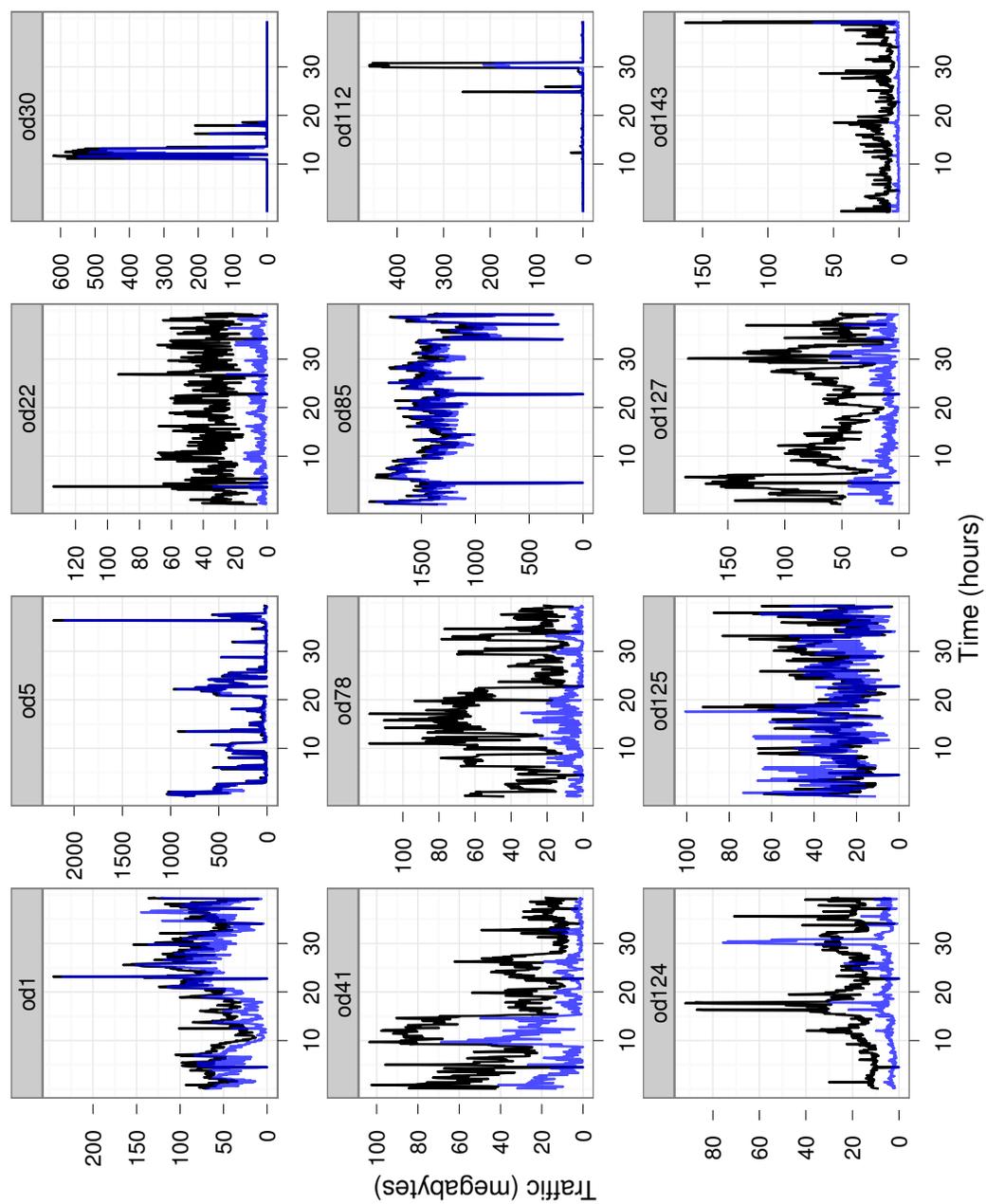

Figure S7: Fitted values vs. ground truth for CMU data. Ground truth in black; Tebaldi & West (joint proposal) in blue.

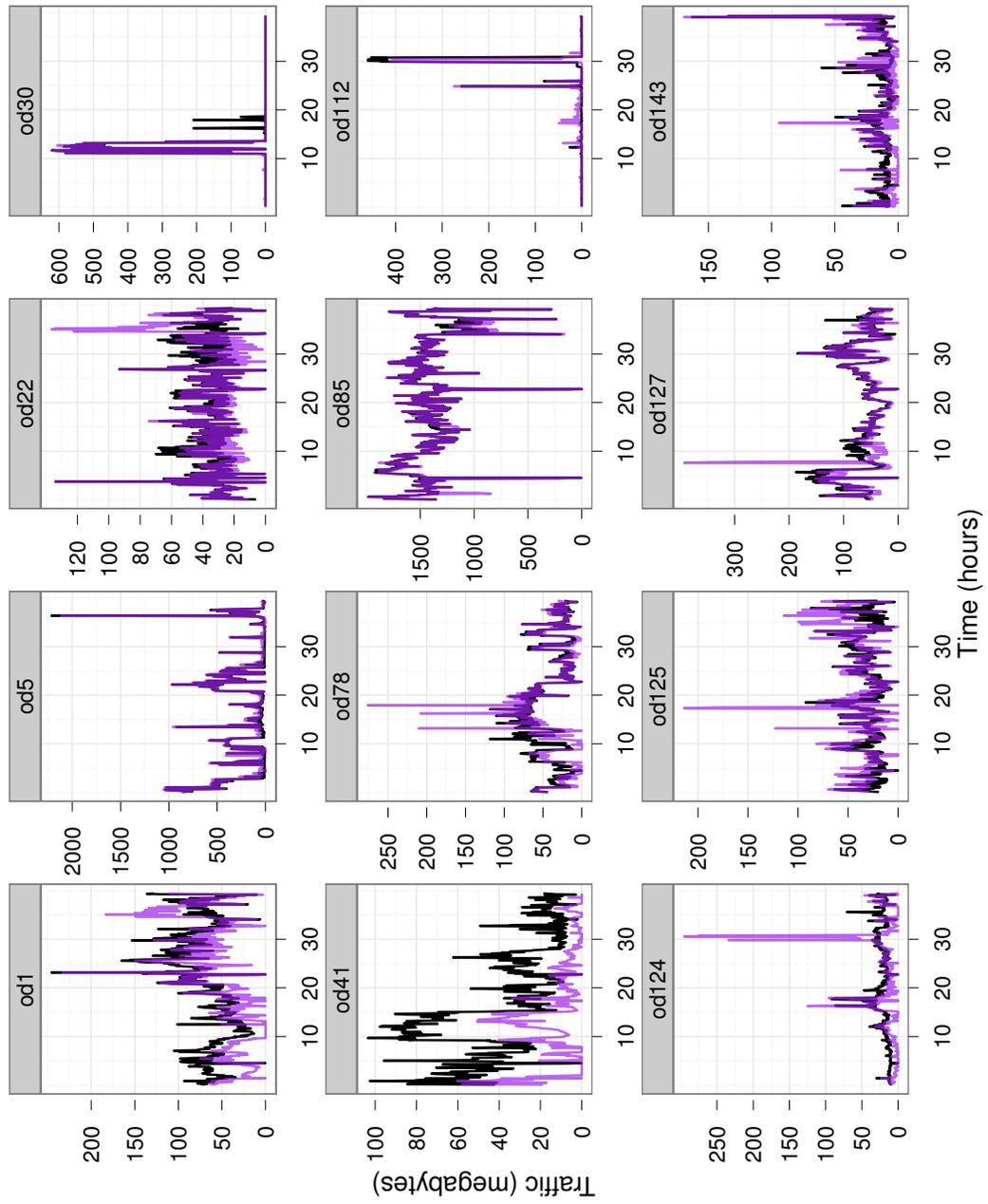

Figure S8: Fitted values vs. ground truth for CMU data. Ground truth in black; Calibration model (stage 1) in purple.

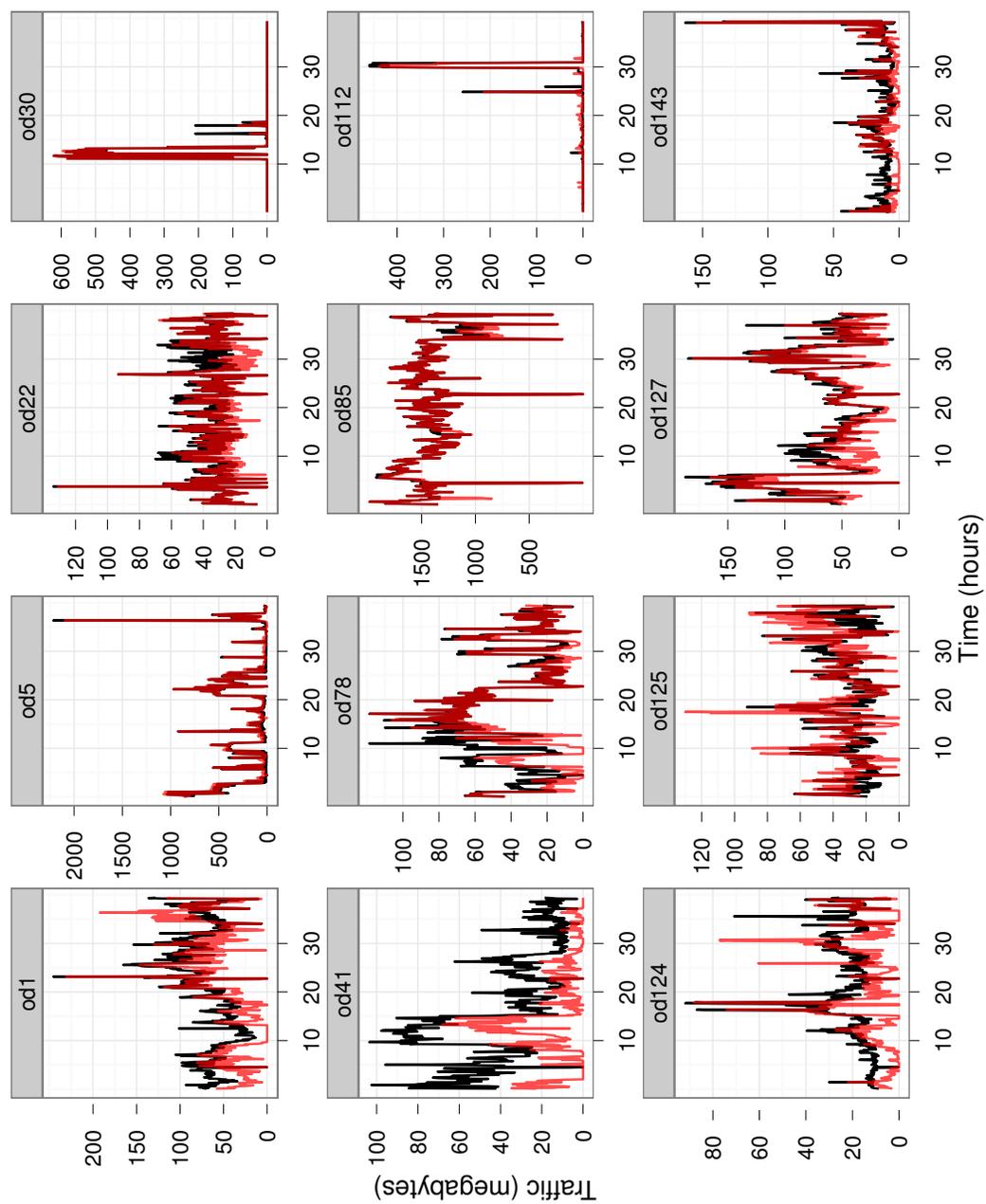

Figure S9: Fitted values vs. ground truth for CMU data. Ground truth in black; Dynamic multilevel model (stage 2) in red.

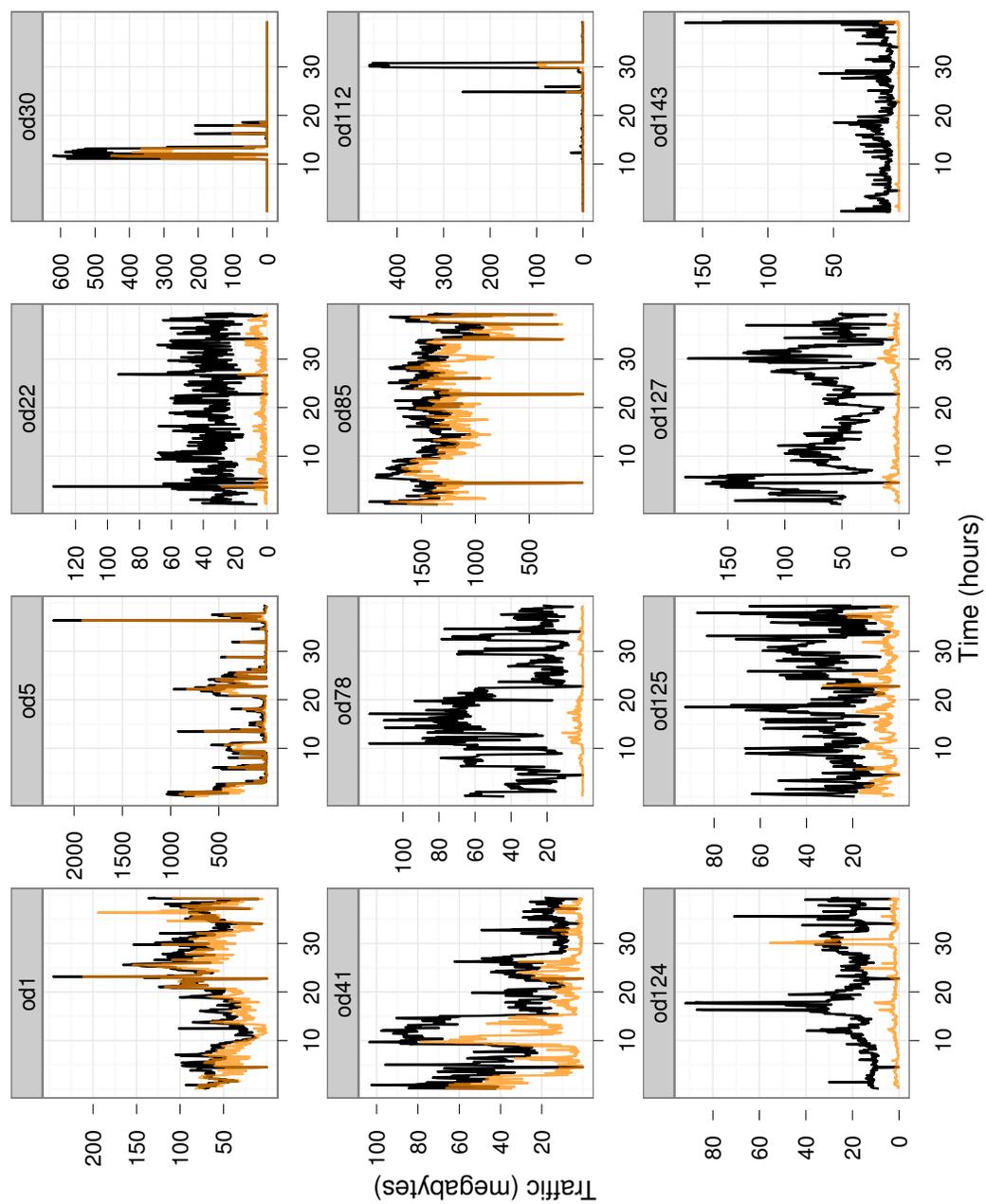

Figure S10: Fitted values vs. ground truth for CMU data. Ground truth in black; Naïve prior in orange.